\documentclass[conference]{IEEEtran}
\makeatletter
\def\ps@headings{%
\def\@oddhead{\mbox{}\scriptsize\rightmark \hfil \thepage}%
\def\@evenhead{\scriptsize\thepage \hfil \leftmark\mbox{}}%
\def\@oddfoot{}%
\def\@evenfoot{}}
\makeatother
\usepackage{amsmath}
\usepackage{amsfonts}
\usepackage{amssymb}
\usepackage{amsthm}
\usepackage{tikz}

\usetikzlibrary{arrows,decorations.pathmorphing,backgrounds,positioning,fit,petri}

\newtheorem{lemma}{Lemma}

\newtheorem{example}{Example}
\newtheorem{theorem}{Theorem}
\newtheorem{corollary}{Corollary}

\begin{document}
%
% paper title
\title{Network Coding Based on Chinese Remainder Theorem}
%
%
% author names and IEEE memberships
% note positions of commas and nonbreaking spaces ( ~ ) LaTeX will not break
% a structure at a ~ so this keeps an author's name from being broken across
% two lines.
% use \thanks{} to gain access to the first footnote area
% a separate \thanks must be used for each paragraph as LaTeX2e's \thanks
% was not built to handle multiple paragraphs
\author{Zhifang~Zhang\\ Key
Laboratory of Mathematics Mechanization, AMSS, Chinese Academy of
Sciences \\E-mail: zfz@amss.ac.cn
%\thanks{Research supported in part by the National Natural
%Science Foundation of China (No.61121062/F02, No.11001254) and the
%Foundation of President of AMSS, CAS.}
}

\maketitle

\begin{abstract}

Random linear network code has to sacrifice part
of bandwidth to transfer the coding vectors, thus a head of size $k\log |T|$ is appended to each packet.
We present a distributed random network coding approach based on the
Chinese remainder theorem for general multicast networks. It uses a couple of modulus
as the head, thus reduces the size of head to $O(\log k)$.
This makes it more suitable for scenarios where the number of source nodes is  large
and the bandwidth is limited. We estimate the multicast rate and show it is satisfactory in performance for randomly designed
networks.

\end{abstract}

\begin{keywords}
Network coding, the Chinese remainder theorem, multi-source
multicast networks
\end{keywords}

\IEEEpeerreviewmaketitle

\section{Introduction}
In their pioneering work, Ahlswede et al \cite{ACLY00:InfoFlow}
state multicast rate can be close to the max-flow bound by
allowing network coding instead of just routing. Then linear coding schemes for general multicast networks are designed \cite{LYC03:linear,Jaggi05:PolyTimeAlgo} which achieve the optimum multicast rate.  Inspired
by these theoretical results, network coding has become a promising
technique to be applied in networking applications, such as wireless
networks and content distribution networks.

In practice, random linear network coding \cite{HKMKE03:RandomNC} is more preferable since it is suitable for dynamic networks. But this approach has to sacrifice part
of bandwidth to keep track of the linear combinations chosen
currently. Namely, each packet needs append a head which is a vector
(called the {\it coding vector}) indicating the combination
coefficients associated with this packet. The overhead of coding
vectors is acceptable for large packets, however, in wireless
applications, such as sensor networks where packets are much shorter
and bandwidth is very limited, it can very fast become prohibitive.
We restate an example described in \cite{Fra10:beyond} and
\cite{SKFAD09:compressed}.
\begin{example}\label{eg1}
Consider a sensor network consisting of $100$ nodes, each sending a
message to a sink. To implement a network code over a field of size
$q=2^4$, the coding vector is in $\mathbb{F}_{2^4}^{100}$ and so is
of $50$ bytes. But in an usual sensor network, such as TinyOs
operating system, a typical frame length allows approximately $30$
bytes for data transmission. Thus just the coding vector alone will
exceed the bandwidth limit.
\end{example}

To shorten coding vectors, paper \cite{SKFAD09:compressed} proposed a
compression approach by constraining the number of nonzero
components of each coding vector no larger than a fraction of the total
dimension. But in practice, to achieve good multicast rate, this
fraction cannot be too small and it is difficult for internal
nodes to maintain this constraint. This approach was later
improved by  using erasure decoding and list
decoding at the cost of increasing decoding complexity at receivers \cite{LR10:improved}.

Another linear coding approach is subspace coding
\cite{KK08:subspaceCod} where messages are mapped into linear subspaces to be transferred and thus no coding vectors are needed. But it
achieves the same information rate as the coding vector based
approach when the packet length increases \cite{Fra10:beyond}.
Moreover, a large codebook must be maintained at the source and sink nodes, and designing subspace codes for
multi-source network coding is very difficult
\cite{SKFAD09:compressed}.

In this paper we propose a distributed random network coding
approach based on the Chinese remainder theorem (CRT) for general
multicast networks. Unlike the random linear network code, it uses a
couple of modulus as the head. The existing
random linear network coding approaches \cite{HKMKE03:RandomNC},
including the compression approach and its improvements
\cite{SKFAD09:compressed,LR10:improved}, all need coding vectors of
size $k\log q$ assuming coding over a field $\mathbb{F}_q$, while
in our coding scheme the counterpart is of size $O(\log k)$.

Before our work, the Chinese remainder theorem has been used in
network coding \cite{CLP08:CRTforward}, but they use CRT-based
coding only at source nodes, and just routing at internal nodes. We
use CRT-based coding at each node, therefore it achieves a higher
multicast rate than just routing. Meanwhile, computation performed at
each node in our scheme can  be simplified by pre-computation.

The paper is organized as follows. Section II introduces the coding vector based approach. Section III describes our CRT-based coding approach for
both single source multicast and multi-source multicast. Section IV gives an elementary estimation of the
multicast rate and displays some experimental results.

\section{Preliminaries}

\subsection{Coding vector based approach for multicast}
Linear network coding is a widely studied approach in the
literature. For convenience, let the alphabet $F$ be an
$l$-dimensional vector space over a finite field $\mathbb{F}_q$,
i.e., $F=\mathbb{F}_q^{\;l}$ for some positive integer $l$, and
$\mathcal{F}$ be a finite dimensional vector space over $F$. A linear network code
can be described as
follows. Let $\mathbf{x}_1,...,\mathbf{x}_k\in\mathbb{F}_q^{\;l}$ be
the information sources. Each receiver $t\in T$ receives
$(\mathbf{s}_1,...,\mathbf{s}_u)$. Since each node performs linear
operations on its input symbols to generate its output symbols,
$\mathbf{s}_i$ is a linear combination of
$\mathbf{x}_1,...,\mathbf{x}_k$. That is, the receiver $t$ gets the
following system of linear equations.
\begin{equation}\label{eq4}
\left(\begin{array}{c}\mathbf{s}_1\\\mathbf{s}_2\\\vdots\\\mathbf{s}_{u}\end{array}\right)=
\underbrace{\left(\begin{array}{cccc}a_{11}&a_{12}&\cdots&a_{1k}\\a_{21}&a_{22}&\cdots&a_{2k}\\\vdots&\vdots&\ddots&\vdots\\
a_{u1}&a_{u2}&\cdots&a_{uk}\end{array}\right)}_
{A\in\mathbb{F}_q^{u\times k}}
\left(\begin{array}{c}\mathbf{x}_1\\\mathbf{x}_2\\\vdots\\\mathbf{x}_{k}\end{array}\right)
\end{equation}
The $i$th row of the coefficient matrix $A$ in (\ref{eq4}) is the
{\it coding vector} corresponding to the received packet
$\mathbf{s}_i$.

Obviously, the receiver needs the knowledge of coding vectors to
solve the equations (\ref{eq4}) and recover the source information
$\mathbf{x}_1,...,\mathbf{x}_k$. In deterministic network coding
\cite{Jaggi05:PolyTimeAlgo,LYC03:linear}, the coding vectors are
pre-determined based on the network topology. For networks with
unknown or changing topologies, distributed random coding
\cite{HKMKE03:RandomNC} is a more useful approach. That is, each
node independently chooses a random linear combination of its
received packets and generates its output packets. Since these
linear combinations are randomly chosen in a distributed setting,
each packet needs append a head to record the coding vector
associated with this packet. Namely, each link $e\in E$ transmits
$$\hat{\mathbf{s}}_e=[\;\mathbf{p}_e\mid\mathbf{s}_e\;]\;,$$
where $\mathbf{s}_e\in\mathbb{F}_q^{\;l}$ is the information packet
and $\mathbf{p}_e\in\mathbb{F}_q^{k}$ is the head such that
$$\mathbf{s}_e=\mathbf{p}_e(\mathbf{x}_1,...,\mathbf{x}_k)^\tau\;.$$
Therefore, in random linear network coding partial bandwidth must
sacrifice to keep track of the coding vectors. In general, it
requires $q>|T|$ to devise a linear network coding over
$\mathbb{F}_q$ \cite{HKMKE03:RandomNC,Jaggi05:PolyTimeAlgo}. When
$k$ and $|T|$ are quite large while the bandwidth is limited, this
coding vector based approach turns out to be constrained.

\subsection{The Chinese remainder theorem}
The Chinese remainder theorem (CRT) is a result about congruences in
number theory. We refer to \cite{Rosen84:NumTheo} for the main
results introduced in this section.

\begin{theorem}(The Chinese Remainder Theorem)

Let $m_1, m_2,...,m_r$ be pairwise relatively prime positive
integers. Then the system of congruences $$x\equiv
a_i\;(\mbox{mod}\; m_i),\;\; 1\leq i\leq r $$has a unique solution
modulo $M=m_1m_2\cdots m_r$.
\end{theorem}

For two integers $a,b$, let $\mbox{gcd}(a,b)$ denote the {\it
greatest common divisor} of $a$ and $b$, and $\mbox{lcm}(a,b)$
denote the {\it least common multiple}. Then we consider systems of
congruences when the modulus are not pairwise relatively prime.

\begin{theorem}\label{th2}
The system of congruences
\begin{eqnarray*}
x&\equiv& a_1\;(\mbox{mod}\; m_1)\\
x&\equiv& a_2\;(\mbox{mod}\; m_2)
\end{eqnarray*}
has a solution if and only if $\mbox{gcd}(m_1,m_2)\mid (a_1-a_2)$.
If there is a solution, it is unique modulo $\mbox{lcm}(m_1,m_2)$.
\end{theorem}

\begin{corollary}
The system of congruences
\begin{equation}\label{eq5}x\equiv a_i\;(\mbox{mod}\; m_i),\;\; 1\leq i\leq r\end{equation}
has a solution if and only if $\mbox{gcd}(m_i,m_j)\mid (a_i-a_j)$
for all pairs of integers $(i,j)$ with $1\leq i<j\leq r$. If there
is a solution, it is unique modulo $\mbox{lcm}(m_1,...,m_r)$.
\end{corollary}

It can use the extended Euclidean algorithm to get a solution to
(\ref{eq5}) in polynomial time.

\section{Network Coding Based on CRT}

\subsection{Single source multicast}\label{sec3.a}
Let $s\in V$ be the source node and $T\subset V$ be the set of
receivers. Denote $|out(s)|=k$ and $|in(t)|=l_t$ for $t\in T$.
Let the
information source be an $n$-bit integer $X$, i.e., $\lceil\log
X\rceil=n$. In the following, we design a network code for the
single-source multicast problem based on the Chinese remainder
theorem.

\noindent{\it Coding at the source}:
\begin{itemize}
\item[(s.1)] The source $s$ randomly selects $2k$ distinct $m$-bit
primes $p_1,...,p_{2k}$. We will discuss the selection of $m$ in Part B.

\item[(s.2)] For $1\leq i\leq k$, the $i$th output link of $s$
transmits $$[X\;\mbox{mod}\;p_{2i-1}p_{2i}\mid p_{2i-1},p_{2i}]\;.$$
\end{itemize}

\noindent{\it Coding at the internal node $v\in V$}:
\begin{itemize}
\item[(v.1)] Suppose the node $v$ gets from all its input links the input $[a_1\mid
q_1,q_2],...,[a_{l_v}\mid q_{2l_v-1},q_{2l_v}]$, where
$l_v=|in(v)|$. Then $v$ solves the system of congruences $$x\equiv
a_i\;(\mbox{mod}\; q_{2i-1}q_{2i}),\;\;1\leq i\leq l_v$$ and gets a
solution $x\equiv a\;(\mbox{mod}\; \mbox{lcm}(q_1,...,q_{2l_v}))$.

\item[(v.2)] $v$ randomly picks $q_i,q_j\in\{q_1,...,q_{2l_v}\}$ and
computes $a'\equiv a\;(\mbox{mod}\;q_iq_j)$. Then its output links
transmit $[a'\mid q_i,q_j]$.
\end{itemize}

\noindent{\it Decoding at the receiver node $t\in T$}:
\begin{itemize}
\item[(t.1)] Suppose the receiver $t$ gets from its input links the
input $[a'_1\mid q'_1,q'_2],...,[a'_{l_t}\mid
q'_{2l_t-1},q'_{2l_t}]$. Then $t$ solves the system of congruences
$$x\equiv a'_i\;(\mbox{mod}\; q'_{2i-1}q'_{2i}),\;\;1\leq
i\leq l_t$$ and gets a solution $x\equiv c\;(\mbox{mod}\;
\mbox{lcm}(q'_1,...,q'_{2l_t}))$.

\item[(t.2)] Denote $\mbox{lcm}(q'_1,...,q'_{2l_t})=N$ and let $0\leq c<N$.
If $\lceil\log (c+N)\rceil>n$, then $t$ recovers $X=c$. Otherwise,
$t$ concludes that $X=c+lN$ for some integer $l$.
\end{itemize}

The following example illustrates our coding approach on the
butterfly network.

\begin{example}
The source $s$ is to multicast the information $X=200$ which is a
$8$-bit integer to the receivers $t_1$ and $t_2$. It chooses primes
$3,5,7,11$ and transmits the information as shown in Figure
\ref{fg1}.

\begin{figure}[ht]
\begin{center}
\begin{tikzpicture}
\tiny
\node [minimum size= 4.5mm,circle,draw] (s) at(-2.5,2.5) {$S$};
\node [minimum size= 4.5mm,circle,draw] (c) at(0.5,2.5) {$c$};
\node [minimum size= 4.5mm,circle,draw] (d) at(2.5,2.5) {$d$};

\node [minimum size= 4.5mm,circle,draw] (a) at(-1,3.5) {$a$};
\node [minimum size= 4.5mm,circle,draw] (b) at(-1,1.5) {$b$};

\node [minimum size= 4.5mm,circle,draw] (t1) at(3.5,3.5) {$t_1$};
\node [minimum size= 4.5mm,circle,draw] (t2) at(3.5,1.5) {$t_2$};

\draw [->] (s) to node [auto] {} (a);
\draw [->] (s) to node [auto] {} (b);
\draw [->] (a) to node [auto] {$[2\mid 3,11]$} (c);
\draw [->] (b) to node [auto] {} (c);
\node at(0.3,1.8) {$[25\mid 5,7]$};
\node at(-1.3,2.8) {$[2\mid 3,11]$};
\node at(-1.3,2.1) {$[25\mid 5,7]$};
\draw [->] (c) to node [auto] {$[46\mid 7,11]$} (d);
\draw [->] (a) to node [auto] {} (t1);
\draw [->] (b) to node [auto] {} (t2);
\draw [->] (d) to node [auto] {} (t1);
\draw [->] (d) to node [auto] {} (t2);
\node at(-2.8,2) {$X=200\in\mathbb{Z}_{256}$};

\end{tikzpicture}
\caption{Source $s$ is to multicast $X=200$ to $t_1$ and
$t_2$}\label{fg1}
\end{center}
\end{figure}
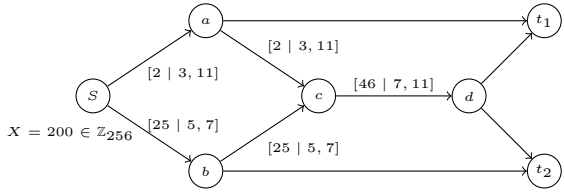

The receiver $t_1$ gets the system of congruences
\begin{eqnarray*}
x&\equiv& 2\;(\mbox{mod}\; 3\times 11)\\
x&\equiv& 46\;(\mbox{mod}\; 7\times 11)
\end{eqnarray*}
By using the extended Euclidean algorithm it obtains a solution
$x\equiv 200\;(\mbox{mod}\; 231)$. Since the information source is
at most $8$-bit in length (which is regarded  as a
predetermined information publicly known to all nodes), $t_1$ can
deduce $X=200$.
\end{example}

\subsection{Parallelization}\label{sec3.b}
Let $(X_1,...,X_u)$ be the information sources where $X_i$ is an
$n$-bit integer. The information is transmitted in the same way as
described in Section \ref{sec3.a}, except that each node deals with
$u$ systems of congruences simultaneously, each for a information
source $X_i$. Thus each packet is of the form
$$[a_1,...,a_u\mid p,q]$$ where $a_i$ is congruent to $X_i$ modulo
$pq$.  Suppose a receiver $t\in T$ gets inputs
$[a'_{1j},...,a'_{uj}\mid q_{2j-1},q_{2j}],1\leq j\leq l_t$. It can
solve the $u$ systems of congruences
$$x_i\equiv a'_{ij}\;(\mbox{mod}\; q_{2j-1}q_{2j}),\;\;1\leq
j\leq l_t,\;\;\;1\leq i\leq u\;,$$ and get the solutions
$$x_i\equiv c_i\;(\mbox{mod}\;\mbox{lcm}(q_1,...,q_{2l_t})),\;\;1\leq i\leq u\;. $$

In practice, the choice of $u$ depends on the
bandwidth and size of the modulus, i.e., the bit
size $m$.  First, determine $m$  according to how many primes are needed in
the network code. For example, in a sensor network with $30$-byte
bandwidth and $100$ source nodes, our CRT-based network code needs
$200$ primes. The {\it prime number theorem} \cite{Ingham90:PNT}
approximates the number of primes no more than  $n$ by
$\frac{n}{\ln n}$. By this approximation, we know the number of
$16$-bit primes is more than $10^3$ which is absolutely enough for
ordinary sensor network. Let $m=16$, then each head message is of
$4$ bytes taking $13.3\%$ of the bandwidth.

Step (t.2) shows there is a possibility that the receiver can only get partial information of the message. Actually fixing $X_i$ to be an integer of size less than $2m$-bit can eliminate this possibility. Then adjusting the value of $u$ according to the bandwidth and size of the message to be transferred.

\subsection{Multi-source multicast}\label{sec3.c}
In sensor networks, the information sources are usually generated at
distributed multiple source nodes and each receiver tries to collect
the information from all sources. In this section we demonstrate how
the CRT-based network code works for multi-source multicast.

Suppose there are $k$ information sources $X_1,...,X_k$ generated at
the distributed source nodes $s_1,...,s_k$ respectively. Let
$T$ be the set of receivers. Each receiver $t\in T$ tries
to get the information $X_1,...,X_k$. Certainly, we assume that
there is a path from source $s_i$ to $t$ for $1\leq i\leq k$ and all
these $k$ paths are edge disjoint.

First, each source node determines a pair of $m$-bit primes
as its identity, making sure that different sources do not have
common primes. That is, let $p_1,...,p_{2k}$ be $m$-bit primes
different from each other, and $(p_{2i-1},p_{2i})$ be $s_i$'s
identity for $1\leq i\leq k$. The identity can be easily determined
in an initial phase. Once they are determined, they are fixed for
all the transmission thereafter and become the common knowledge to
all nodes.

Without loss of generality, we assume $X_i$ is a $(2m-1)$-bit
integer since integers of larger size can be cut in parts and
transmitted in parallel. Then for $1\leq i\leq k$ the source $s_i$
sends $[X_i\mid p_{2i-1}, p_{2i}]$ on each of its output links.
In this way the $k$ sources jointly determine an integer
$X$ satisfying the system of congruences:
$$X\equiv X_i\;(\mbox{mod}\; p_{2i-1} p_{2i}),\;\;1\leq i\leq k\;.$$
From Theorem \ref{th2} we know this system is solvable although the
equations are determined in a distributed setting.

The internal nodes do the same as in steps (v.1) and (v.2). Then for
a receiver $t\in T$ with $l_t$ input links, it solves the system of
congruences as in step (t.1) and gets a solution $x\equiv
c\;(\mbox{mod}\; \mbox{lcm}(q'_1,...,q'_{2l_t}))$. It is easy to see
that $X\equiv c\;(\mbox{mod}\; \mbox{lcm}(q'_1,...,q'_{2l_t}))$ and
$\{q'_1,...,q'_{2l_t}\}\subseteq\{p_1,...,p_{2k}\}$. There are three
cases for the receiver $t$ recovering $X_i$:
\begin{itemize}
\item[(1)] If $\{p_{2i-1},p_{2i}\}\subseteq\{q'_1,...,q'_{2l_t}\}$,
then $t$ can recover $X_i$ as $X_i\equiv c\;(\mbox{mod}\;
p_{2i-1}p_{2i})\;$.

\item[(2)] If
$\{p_{2i-1},p_{2i}\}\cap\{q'_1,...,q'_{2l_t}\}=\{p_{2i}\}$, then $t$
computes $c'\equiv c\;(\mbox{mod}\;p_{2i})$ and concludes
$X_i=c'+bp_{2i}$ for some integer $b$. A similar conclusion can be
made when
$\{p_{2i-1},p_{2i}\}\cap\{q'_1,...,q'_{2l_t}\}=\{p_{2i-1}\}$.

\item[(3)] If
$\{p_{2i-1},p_{2i}\}\cap\{q'_1,...,q'_{2l_t}\}=\emptyset$, then $t$
cannot recover $X_i$.
\end{itemize}

Although  there is a chance that $t$ cannot recover $X_i$ or just
know partial information about $X_i$, we will show in section
\ref{sec4} that this chance can be very small when $l_t$ is large
enough.

\subsection{Simplifying computation}
In steps (v.1) and (v.2) of the CRT-based network coding, the
internal node with $l_v$ input links needs to solve a system of
$l_v$ congruent equations  and then pick two primes as its output
modulus. Actually, this process can be simplified. The internal node
first picks two primes from all his input primes, say, $p$ and $q$.
Let $[a_i\mid p,p']$ and $[a_j\mid q,q']$ be two packets it received
containing the picked primes $p$ and $q$. Compute $$a\equiv
a_i\;(\mbox{mod}\; p)\;\;\mbox{and}\;\;b\equiv a_j\;(\mbox{mod}\;
q)\;.$$ Then, it has a system of two congruences:
\begin{equation}\label{eq3}X\equiv a\;(\mbox{mod}\;
p)\;\;\mbox{and}\;\;X\equiv b\;(\mbox{mod}\; q)\;.\end{equation} It
is much easier to solve the system (\ref{eq3}) containing only two
congruent equations and get a solution of $X$ modulo $pq$. Similar
simplifications can be made at the decoding steps in multi-source
case.

\section{Estimation of multicast rate}\label{sec4}

For the single source multicast
, a receiver can always recover the message as stated in Section III-B.
For the multi-source case, a receiver $t$ finally gets modulus $\mbox{lcm}(q_1',...,q_{2l_t}')$, then he can recover $X_i$ if
$\{p_{2i-1},p_{2i}\}\subseteq\{q_1',...,q_{2l_t}'\}$. We are
interested in the number of $X_i$'s that a receiver can recover.

\begin{lemma}\label{le2}
Let $k,l$ be positive integers. Let $S_i=\{2i-1,2i\}$ for $1\leq
i\leq k$ and $S=\cup_{i=1}^kS_i$. For $1\leq j\leq l$, a subset
$A_j\subseteq \mathcal{S}$ with $|A_j|=2$ is independently and
uniformly chosen. Denote $$I_S=\{i\mid 1\leq i\leq k,\;
S_i\subseteq\cup_{j=1}^lA_j\}\;.$$ Then the expectation of $|I_S|$
is $k(1-(1-\frac{1}{k})^l)^2$.
\end{lemma}

Denote $l=rk$. Since
$\lim\limits_{k\to\infty}(1-\frac{1}{k})^k=\frac{1}{e}\approx0.3679$,
we approximate $(1-\frac{1}{k})^k$ by $0.367$ as $k$ is a
three-digit number.
Then the expectation of $|I_S|$ is $k(1-0.367^r)^2$. Define the
recover rate as $R^*=|I_S|/k$. Thus the expectation of $R^*$ is
about $(1-0.367^r)^2$. Since a receiver tries to collect all source
information, it is reasonable to assume that for any $t\in T$,
$l_t\geq k$, i.e., $r\geq 1$. Table \ref{ta2} lists our estimation
of $R^*$ at some points of $r$. It can see as $r$ increases the
recover rate $R^*$ becomes more and more satisfactory.

\begin{table}[ht]
$$\begin{array}{|c|c|c|c|c|c|c|c|}\hline
r&1&1.5&2&2.5&3&3.5&4\\\hline
\/R^*\/&\/0.40\/&\/0.60\/&\/0.75\/&\;0.84\/&\/0.90\/&\/0.94\/&\/0.96\/\\\hline\end{array}$$
\caption{}\label{ta2}
\end{table}

\subsection{An experiment}

Our estimation  is based on the probabilistic
event described in  Lemma \ref{le2}.  In
the following, we show this estimation does
not deviate the real performance too much for a randomly designed network.

The network is designed as follows. First, all nodes in the network
are divided into $L+2$ levels, denoted as
$V_0,V_1,...,V_L,V_{L+1}\subset V$, and each level contains $M_i$
nodes, i.e., $|V_i|=M_i$. Let $V_0$ be the set of sources and
$V_{L+1}$ be the set of receivers, and the rest be internal nodes.
For $0\leq i\leq L$, each node in $V_i$ independently links to
$\sigma M_{i+1}$ nodes which are randomly and uniformly chosen in
$V_{i+1}$.

In a experiment, set $M_0=100, M_{L+1}=10, \sigma=0.8$, and
$M_1=\cdots=M_L=M$. Then implement the CRT-based network code on
such a network at different values of $M$ and $L$, and record the
number of primes each receiver finally gets. The results is
displayed in Table \ref{ta3}, where $t_i$ means the number of primes that the $i$th
receiver finally collects and $\mathcal{R}'=\frac{\frac1{10}\sum_{i=1}^{10}t_i}{200}$ which is close to the real recover rate. Note in the experiment $r\approx\frac{\sigma M}{M_0}$. It can see when $r$ increases, the recover rate improves evidently. Meanwhile, each link only takes a fraction (for example $13.3\%$ as described in Section III-B) of its bandwidth to transfer the head message, which means the multicast rate is close to the optimal rate determined by the max-flow bound.

\begin{table}[ht]
$$\begin{array}{|c|c|c|c|c|c|c|c|c|c|c|c|}\hline
M,L\!&t_1&t_2&t_3&t_4&t_5&t_6&t_7&t_8&t_9&t_{10}&\mathcal{R}'\\\hline
\!200,5\!&\!155\!&\!157\!&\!160\!&\!160\!&\!156\!&\!156\!&\!161\!&\!157\!&\!154\!&\!157\!&\!0.787\!\\\hline
\!200,3\!&\!159\!&\!155\!&\!156\!&\!161\!&\!156\!&\!159\!&\!156\!&\!158\!&\!154\!&\!155\!&\!0.785\!\\\hline
\!250,5\!&\!173\!&\!169\!&\!173\!&\!166\!&\!175\!&\!175\!&\!171\!&\!175\!&\!170\!&\!175\!&\!0.861\!\\\hline
\!250,3\!&\!172\!&\!171\!&\!175\!&\!167\!&\!166\!&\!170\!&\!169\!&\!172\!&\!174\!&\!169\!&\!0.853\!\\\hline
\!400,5\!&\!195\!&\!192\!&\!194\!&\!194\!&\!192\!&\!191\!&\!189\!&\!194\!&\!194\!&\!194\!&\!0.965\!\\\hline
\!400,3\!&\!187\!&\!195\!&\!193\!&\!195\!&\!194\!&\!195\!&\!192\!&\!192\!&\!190\!&\!195\!&\!0.963\!\\\hline
\end{array}$$
\caption{}\label{ta3}
\end{table}

Actually it is difficult to achieve the optimal rate for all distributed random network coding, since partial bandwidth is taken by the head message.
The main advantage of our CRT-based network coding is the great
reduction in head message. Suppose there are $k$ source nodes
or the information source is of $k$ dimensions, and there are $|T|$
receivers. Then the coding vector based network code needs to convey
the head message which is a $k$-dimensional vector over
$\mathbb{F}_q$, where $q>|T|$. Thus the head message is of size
$k\log |T|$. Sometimes the size of this head message alone will exceed the bandwidth. While our CRT-based network code only needs to convey a
pair of primes chosen from $2k$ distinct $m$-bit primes. By
the prime number theorem, the primes are of size $O(\log k)$.
Therefore, comparing with the coding vector based approach, we
reduce the size of head message from $k\log |T|$ to $O(\log
k)$.

% biography section
%
% If you have an EPS/PDF photo (graphicx package needed) extra braces are
% needed around the contents of the optional argument to biography to prevent
% the LaTeX parser from getting confused when it sees the complicated
% \includegraphics command within an optional argument. (You could create
% your own custom macro containing the \includegraphics command to make things
% simpler here.)
%\begin{biography}[{\includegraphics[width=1in,height=1.25in,clip,keepaspectratio]{mshell}}]{Michael Shell}
% where an .eps filename suffix will be assumed under latex, and a .pdf suffix
% will be assumed for pdflatex; or if you just want to reserve a space for
% a photo:

%\begin{biographynophoto}{Zhifang Zhang}
%received the PhD degree in applied mathematics from the Academy of
%Mathematics and Systems Science, the Chinese Academy of Sciences, in
%2007.
%\end{biographynophoto}

\vfill
% You can push biographies down or up by placing
% a \vfill before or after them. The appropriate
% use of \vfill depends on what kind of text is
% on the last page and whether or not the columns
% are being equalized.

%\vfill

% Can be used to pull up biographies so that the bottom of the last one
% is flush with the other column.
%\enlargethispage{-5in}

% that's all folks
\end{document}